# A Recurrent Nova Super-Remnant in the Andromeda Galaxy

M. J. Darnley[1], R. Hounsell[2,3], T. J. O'Brien[4], P. Rodríguez-Gil[5,6],

A. W. Shafter[7], M. M. Shara[8], M. Henze[7], M. F. Bode[1,9],

R. Galera-Rosillo[5,6], D. J. Harman[1], J.-U. Ness[10], V. A. R. M. Ribeiro[11,12],

N. M. H. Vaytet[13], S. C. Williams[14]

[1]Astrophysics Research Institute, Liverpool John Moores University, IC2 Liverpool Science Park, Liverpool, L3 5RF, UK.

[2]Department of Astronomy and Astrophysics, University of California, Santa Cruz, CA 95064, USA.

[3]Astronomy Department, University of Illinois at Urbana-Champaign, 1002 W. Green Street, Urbana, IL 61801, USA.

[4]Jodrell Bank Centre for Astrophysics, Alan Turing Building, University of Manchester, Manchester, M13 9PL, UK.

[5]Instituto de Astrofísica de Canarias, Vía Láctea, s/n, La Laguna, E-38205, Santa Cruz de Tenerife, Spain.

[6]Departamento de Astrofísica, Universidad de La Laguna, La Laguna, E-38206, Santa Cruz de Tenerife, Spain.

[7]Department of Astronomy, San Diego State University, San Diego, CA 92182, USA.

[8]American Museum of Natural History, 79th Street and Central Park West, New York, NY 10024, USA.

[9]Vice Chancellor's Office, Botswana International University of Science and Technology, Private Bag 16, Palapye, Botswana.

[10]XMM-Newton Observatory SOC, European Space Astronomy Centre, Camino Bajo del Castillo s/n, Urb. Villafranca del Castillo, 28692 Villanueva de la Cañada, Madrid, Spain.
[11]CIDMA, Departamento de Física, Universidade de Aveiro, Campus Universitário de Santiago, 3810-193 Aveiro, Portugal.
[12]Instituto de Telecomunicações, Campus Universitário de Santiago, 3810-193 Aveiro, Portugal.
[13]Centre for Star and Planet Formation, Niels Bohr Institute and Natural History Museum of Denmark, University of Copenhagen, Øster Voldgade 5-7, DK-1350 Copenhagen K, Denmark.
[14] Physics Department, Lancaster University, Lancaster, LA1 4YB, UK.

**The accretion of hydrogen onto a white dwarf star ignites a thermonuclear runaway in the accumulated envelope, leading to luminosities up to 1 million times that of the Sun and a high velocity mass ejection that produces a remnant shell – a classical nova eruption[1,2]. Close to the upper mass limit of a white dwarf[3] (1.4 Msun), rapid accretion of hydrogen (~$10^{-7}$ Msun/yr) from a binary star companion leads to frequent eruptions on timescales of years[4,5] to decades[6]. Such systems are known as recurrent novae. The ejecta of recurrent novae, initially moving at up to 10,000 km/s[(7)], must sweep-up the surrounding interstellar medium and evacuate cavities around the nova binary. No remnant larger than one parsec from any single classical or recurrent nova eruption is known[8,9,10], but thousands of successive recurrent nova eruptions should be capable of generating shells ~100-1,000 times this size. Here we report that the most rapidly recurring nova, M31N 2008-12a, which erupts annually[11], is surrounded by such a nova super-remnant with a projected size of at least 134**

**by 90 parsecs. Larger than almost all known remnants of supernova explosions[12], this enormous ring demonstrates that M31N 2008-12a has erupted with high frequency for millions of years.**

**Subject terms: Stars Novae**

Located within the disk of the Andromeda Galaxy (M31), the rapidly recurring nova M31N 2008-12a has erupted annually since at least 2008[11]. The eruptions of M31N 2008-12a (hereafter '12a') exhibit the fastest optical evolution, the highest ejection velocities, the hottest X-ray source, and the most rapid recurrence cycle of any known thermonuclear nova[11]. Combined, these observations require the most massive white dwarf ever discovered[13] (1.38 Msun), accreting at the largest rate seen in any nova system[14] ($>10^{-7}$ Msun/yr). *Hubble Space Telescope* ultraviolet spectroscopy of the 2015 eruption uncovered no evidence for neon in the ejecta, which is consistent with – but not conclusive proof of – a carbon-oxygen white dwarf[15], one which must have grown from a formation mass of at most 1.1 Msun[5,16,17].

Ground-based narrow-band H$\alpha$+[N II] imaging shows a ring-like structure spatially coincident with the nova[18,19]. The ring of this shell-like nebula surrounding 12a is clearly visible in new deep ground-based and *Hubble Space Telescope* observations (Fig. 1; see Methods) – the proposed nova super-remnant. The super-remnant is elliptical and brighter to the southwest than it is to the northeast. Using these data, we measure the projected semi-major axes to the inner and outer edge of the bright super-remnant shell to be 52 and 67 parsecs, respectively, giving a shell thickness of 22%. There is a sharply defined outer edge clearly visible to the south and the west. The well-defined elliptical boundary of the super-remnant implies that it has not been substantially shaped by the interstellar medium (ISM), but that such a geometry was imparted by the nova eruptions, and has largely persisted. The high spatial resolution of the *Hubble Space Telescope* images reveals that the super-remnant outer shell is not smooth, as seen from the ground, but fragmented into knots and filaments, reminiscent of the handful of interacting nova shells seen around the Galactic recurrent nova T Pyxidis[20,21].

A spectrum of the super-remnant shell (see Fig. 2 and methods) reveals strong and narrow emission lines of the hydrogen Balmer series with widths narrower than the instrumental resolution (~250 km/s for H$\alpha$). The only helium line detected is the He I (587.6nm) recombination line. The presence of the [O II] and [S II] doublets places an upper limit on the electron density of the emitting gas of ~3,000cm$^{-3}$ (ref. 22), and the lack of [O III] indicates that the level of ionisation in the outer shell is low. As the [N II] (654.8/658.4nm) doublet is visible, but the [N II] (575.5nm) line is not, we can place a limit on the electron temperature of <9,000K[(22)]. The [N II]/H$\alpha$ line-intensity ratio is 0.54±0.02, the [S II]/H$\alpha$ ratio is 0.32±0.01, while the [S II] doublet ratio itself is 1.42±0.05 and indicates an electron density <100cm$^{-3}$ (ref. 22) within the bright outer shell of the super-remnant. This density and measurements of the super-remnant shell size (Fig. 1) indicate the shell mass is $<7\times10^5$ Msun (see methods).

The [S II] and [N II] line strengths are intermediate between those expected for planetary nebulae, the remnants of supernova explosions, and H II regions[23], but the lack of [O III] emission indicates there is no nearby source of ionising radiation (such as OB stars) and that the outer shell is of sufficient age to have cooled below the ionisation temperature of $O^+$. A further spectrum contains emission from a bright knot well within the super-remnant and to the east of 12a (see Fig. 1b and methods). This knot spectrum contains [O III] emission, which indicates a more extreme temperature or ultraviolet radiation environment closer to the nova system.

To explore the viability of multiple recurrent nova eruptions producing such a vast super-remnant, we performed a series of one-dimensional hydrodynamic simulations of the ejecta, their self-interaction, and their interaction with the surrounding environment. Results of our simulations of up to 5,000 separate but interacting ejecta are presented in Fig. 3. The simulations (Fig. 3a & 3b) illustrate how repeated nova eruptions create a vast, evacuated cavity around the system, by continually sweeping up the ISM and piling it up within a shell at the edge of the growing super-remnant (Fig. 3d). Assuming spherical geometry, such repeated eruptions sweep up a tenth of a solar mass of ISM after 5,000 eruptions (Fig. 3e), which equates to ~300 times the mass ejected by the nova over this period. Therefore, such super-remnants comprise almost exclusively swept up material and will be expected to have an ISM-like, not nova-like, composition, which is consistent with the super-remnant He I/H$\alpha$ line-strength ratio of 0.021±0.006. Similar ratios are observed in warm diffuse ISM[24]. The He I/H$\alpha$ ratio has been repeatedly measured in the first four days after eruptions of 12a and varies from 0.16-0.48[11], driven by the high abundances of helium in nova ejecta[4].

The super-remnant contains three distinct regions, as marked in Fig. 3b (also see methods): the *inner cavity*, where recent ejecta effectively undergo free, high velocity expansion while cooling adiabatically; the *ejecta pile-up*, where the ejecta from successive eruptions eventually collide as they are slowed by interaction with the ISM, with inter-ejecta shocks driving significant heating of this gas; and the *super-remnant shell*, which consists almost entirely of swept-up ISM that is slowly driven outward by the multiple-ejecta pile-up occurring at its inner edge. The radius of the outer edge of the super-remnant increases with a power-law-like dependence upon time (Fig. 3d), with the inner radius growing more slowly as more and more material is accumulated, maintaining a shell thickness of the outer ~20% of the super-remnant, which is consistent with the observations. The simulations reveal that, once established, the shell over-density remains at a factor of four times the ISM density (Fig. 3a).

The computational intensity limited the simulations to 5,000 eruptions, therefore the continued growth of the super-remnant must be explored by extrapolating the results to larger timescales (see methods). Our models show that a super-remnant the size of that observed can be built up by annual recurrent nova eruptions sweeping up the surrounding ISM over the course of $6\times10^6$ years (Fig. 3d). By that time, the outer shell of the super-remnant will have cooled sufficiently (to well below $10^4$K) to explain the observed spectrum (Fig. 3f). In those $6\times10^6$ years the simulations show that the total mass swept up by the eruptions is $\sim3\times10^4$ Msun, consistent with the upper limit derived from the observations. The size and mass of this super-remnant demonstrate that 12a has not just been erupting frequently for a decade as observed, but for millions of years.

The 12a white dwarf has a mass accretion rate of $\sim1.6\times10^{-7}$ Msun/yr[(13)]. Thus, an average accretion efficiency (the proportion of accreted material retained by the white dwarf after an eruption) of 40% over the proposed age of the super-remnant ($6\times10^6$ years) is required to grow a carbon-oxygen white dwarf from a zero-age mass of ~1 Msun to the maximum mass permissible for a white dwarf before collapse ensues[3] (1.4 Msun; the Chandrasekhar mass). The current accretion efficiency in 12a is >60%[13,14], consistent with the prediction of increasing accretion efficiency as the white dwarf mass grows[5].

The discovery of more nova super-remnants around other accreting white dwarfs will provide additional striking signposts to systems undergoing regular eruptions over a long-period of time. Our simulations show that this super-remnant, the first discovered extragalactic nova shell, is not static and will continue to grow at least as long as nova eruptions continue in the system. As such, any nova super-remnants around accreting carbon-oxygen white dwarfs will ultimately be destroyed by the explosion of their parent nova system as a Type Ia supernova. M31N 2008-12a is predicted to surpass the Chandrasekhar mass limit in less than 40,000 years[14]. At such time, the underlying composition of the white dwarf[15] will be revealed incontrovertibly when either a Type Ia supernova explosion[25], or an accretion induced collapse of the white dwarf to a neutron star[26], is observed.

**Main Figure Legends**

**Figure 1. (a)** Liverpool Telescope narrow-band Hα+[N II] continuum-subtracted (see methods) image of the region surrounding M31N 2008-12a. The majority of stellar sources have been removed, but the eight dark-blue sources indicate field stars only detected in continuum light. The presence of the closed ring nebula is seen within the white dashed ellipse, as is its asymmetry and varying luminosity around the outer 'shell'. The position of M31N 2008-12a is marked and the offset from the geometric centre is indicated by the black line. **(b)** *Hubble Space Telescope* Hα+[N II] continuum-subtracted (see methods) image of the same region; all stellar sources have been removed via the subtraction process. The high spatial resolution of this image reveals that the nebulosity is not smooth as imaged from the ground, but fragmented and filamentary in nature, reminiscent of the ejecta of the Galactic recurrent nova T Pyxidis[21]. The red squares mark the location of the two regions discussed in the text, the large square shows the bright western shell, and the small square the eastern 'knot'. **(c)** Zoomed in *Hubble Space Telescope* Hα+[N II] image showing the region within the large red box in the centre panel. To the top of this panel three long 'nested' filaments are discernible, separated by only 5 and 12 parsecs, respectively.

**Figure 2.** The Gran Telescopio Canarias spectrum of the bright western part of the nova super-remnant shell shown in Fig. 1. This spectrum shows negligible continuum emission punctuated by strong hydrogen Balmer series lines (Hα through Hδ), a He I

recombination line, and forbidden or 'nebular' lines of [N II], [O II], and [S II]. With a spectral resolution of 0.53nm, the H$\alpha$+[N II] and [S II] lines are easily resolved, but the [O II] doublet is blended. No other lines can be reliably confirmed. Gaps in the spectrum indicate areas where significant skyline subtraction residuals remained.

**Figure 3.** Results of the hydrodynamic simulations of the interacting ejecta of multiple recurrent nova eruptions. **(a)** The radial density profile around M31N 2008-12a. The magenta, cyan, blue, green, red, and black lines illustrate the simulated density profile following 2, 10, 20, 40, 70, and 100 eruptions, respectively. The lower black dotted horizontal line indicates the interstellar medium density, with the upper dotted line showing the consistent peak density of the super-remnant shell. **(b)** As (a) but here the solid lines represent 20, 100, 200, 400, 700, and 1,000 eruptions, respectively. **(c)** As (a) and (b) but the solid lines show 1,000, 2,000, and 5,000 eruptions, respectively. **(d)** The upper black solid line illustrates the growth of the outer edge of the super-remnant shell over 5,000 eruptions (spatial resolution 1AU), the lower black solid line shows the position of the inner edge. The blue, red, and green solid lines indicate similar simulations of 1,000, 100, and 20 eruptions with spatial resolutions of 0.4, 0.2, and 0.02AU, respectively. The black dotted lines illustrate extrapolations of the radial growth curves to more eruptions. The upper/lower solid grey lines indicate the growth of the outer edge for lower/higher interstellar medium densities, respectively. The horizontal dotted line marks the observed maximum projected radius of the M31N 2008-12a super-remnant (67 parsecs). **(e)** As (d) but here the super-remnant outer shell mass is shown. The upper/lower solid grey lines show the effect of a higher/lower interstellar medium density. **(f)** The evolution of the electron temperature within the outer shell of the super-

remnant. The solid black line indicates simulations of 5,000 eruptions with resolution of 1AU, the red/green lines show the effects of a lower/higher interstellar medium density, respectively. An extrapolation to more eruptions is again shown by the diagonal black dotted line, while the horizontal dotted line indicates the shell electron temperature upper limit as required by the spectroscopy.

**Methods Figure Legends**

**Methods Figure 1.**

The Steward 2.3m Bok Telescope H$\alpha$ image that allowed the association between the nebulosity and M31N 2008-12a to be made. Image orientation is as Fig. 1 but the image shows four times the field of view.

**Methods Figure 2.**

*Hubble Space Telescope* Wide Field Camera 3 broad-band filter images of the region around M31N 2008-12a. The three panels show the **(a)** F275W (ultraviolet), **(b)** F475W (optical), and **(c)** F814W (optical) filters. The position of M31N 2008-12a is marked by the red circle in the centre of these images and the images are to the same scale as Fig. 1a. The white contours in the centre panel show iso-flux regions as derived from the ground-based H$\alpha$+[N II] image. As these images were taken during the 2015 eruption, the nova can be seen in the images.

**Methods Figure 3.**

The Gran Telescopio Canarias spectrum of the inner eastern knot within the super-remnant shown in Fig. 1. As with the spectrum of the super-remnant shell (Fig. 2), there is negligible continuum and hydrogen Balmer emission lines, but the nebular lines of [N II], [O II], and [S II] have been joined by [O III], indicative of higher excitation. Gaps in the spectrum indicate areas where significant skyline subtraction residuals remained

**Methods Figure 4.**

Comparison of the results from the hydrodynamic modelling using a range of spatial resolutions. The blue and green lines indicate simulations of 20 eruptions with spatial resolutions of 0.02 and 0.2AU, respectively, while the red and black lines indicate simulations of 100 eruptions with resolution 0.2 and 0.4 AU, respectively. (a) Gas density radial distribution, the lower black dotted horizontal line indicates the interstellar medium density, with the upper dotted line showing the consistent peak density of the super-remnant shell. (b) Gas pressure radial distribution. (c) Gas velocity radial distribution. (d) Gas temperature radial distribution.

**Methods Figure 5.**

Panels as Methods Figure 4. A comparison between the results of simulations of 1,000 eruptions without radiative cooling (black) and with radiative cooling (blue).

**Acknowledgments:**

Based on observations made with the NASA/ESA *Hubble Space Telescope*, obtained from the Data Archive at the Space Telescope Science Institute (STScI), which is operated by the Association of Universities for Research in Astronomy, Inc., under NASA contract NAS 5-26555. These observations are associated with programmes #14125 and #14651. Support for programmes #14124 and #14651 was provided by NASA through grants from STScI. RH, MMS, and AWS acknowledge financial support from those grants.

The Liverpool Telescope is operated on the island of La Palma by Liverpool John Moores University (LJMU) in the Spanish Observatorio del Roque de los Muchachos of the Instituto de Astrofísica de Canarias with financial support from STFC.

Based on observations made with the Gran Telescopio Canarias (GTC), installed in the Spanish Observatorio del Roque de los Muchachos of the Instituto de Astrofísica de Canarias, in the island of La Palma.

IRAF is distributed by the National Optical Astronomy Observatory, which is operated by the Association of Universities for Research in Astronomy (AURA) under a cooperative agreement with the National Science Foundation.

The Starlink software is currently supported by the East Asian Observatory.

The authors would like to thank Zoltan Levay, STScI, for assistance creating a colour composite image of the nova super remnant.

MJD thanks M. W. Healy and I. A. Steele for proofreading drafts of this manuscript.

MJD and RH thank M. Link and C. Proffitt, and W. Eck and K. Long, the programme coordinators and contact scientists for *Hubble Space Telescope* programmes #14124 and #14651, respectively, for their assistance with the observations.

AWS thanks K. A. Misselt and D. Baer for their assistance with the Steward 2.3m Telescope observations and data reduction.

VARMR acknowledges financial support from Fundação para a Ciência e a Tecnologia (FCT) in the form of an exploratory project of reference IF/00498/2015, from the Center for Research & Development in Mathematics and Applications (CIDMA) strategic project UID/MAT/04106/2013, and from Enabling Green E-science for the Square Kilometer Array Research Infrastructure (ENGAGE SKA), POCI-01-0145-FEDER- 022217, funded by Programa Operacional Competitividade e Internacionalização (COMPETE 2020) and FCT, Portugal.

NMHV acknowledges support from the European Commission through the Horizon 2020 Marie Sklodowska-Curie Actions Individual Fellowship 2014 programme (Grant Agreement no. 659706).

**Author contributions:**

All authors contributed to the discussion, proposing and planning of observations, data interpretation, and writing of this manuscript. AWS acquired the Steward 2.3m Bok Telescope data that was subsequently used to discover the nebulosity surrounding the system. MJD and SCW led the Liverpool Telescope imaging and observations. MJD and RH led the Hubble Space Telescope observations of the nova super-remnant. PR-G and MJD led the Gran Telescopio Canarias observations, RG-R assisted with their analysis. TJO'B and MJD led the hydrodynamic simulations of the nova eruptions. TJO'B and NMHV developed and supported the use of the Morpheus hydrodynamic simulation code.

**Competing Financial Interests**

The authors declare no competing financial interests.

**Correspondence Author**

Correspondence and requests for materials should be addressed should be addressed to MJD (M.J.Darnley@ljmu.ac.uk).

**Data Availability**

All relevant data are available from the corresponding author on reasonable request.

**Figure 1.**

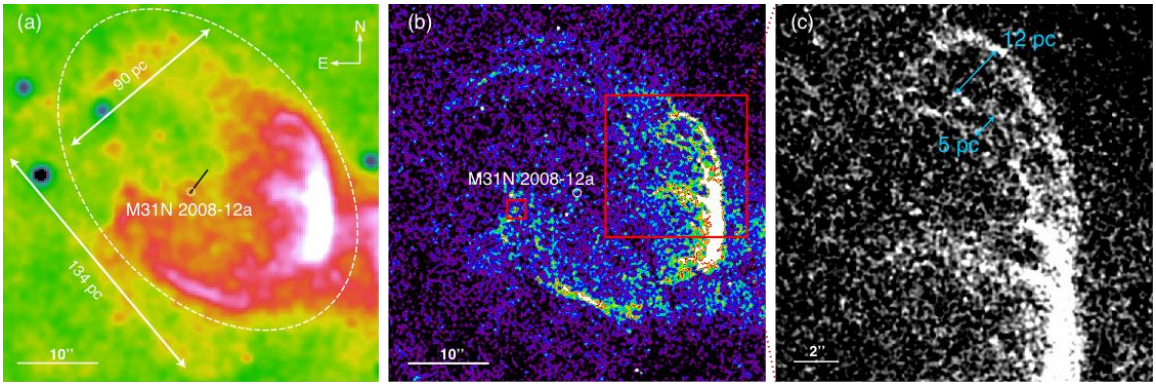

**Figure 2.**

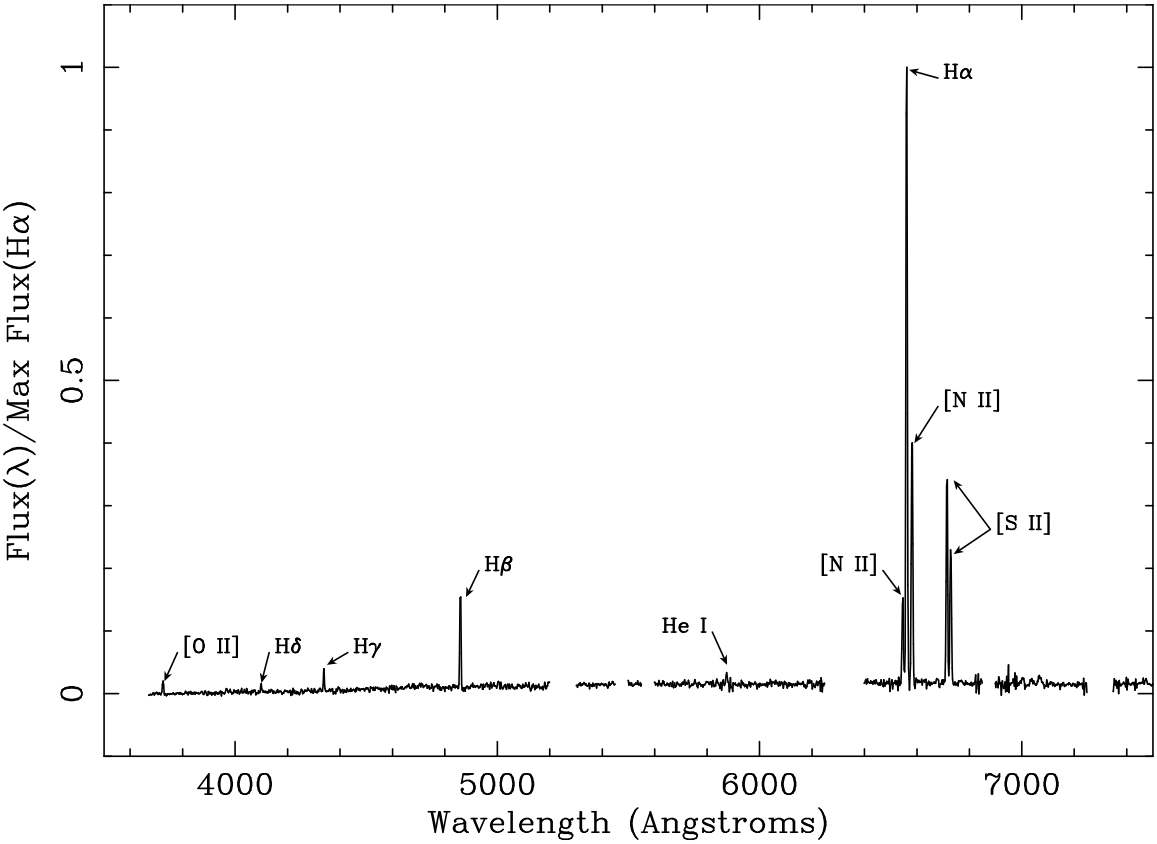

**Figure 3.**

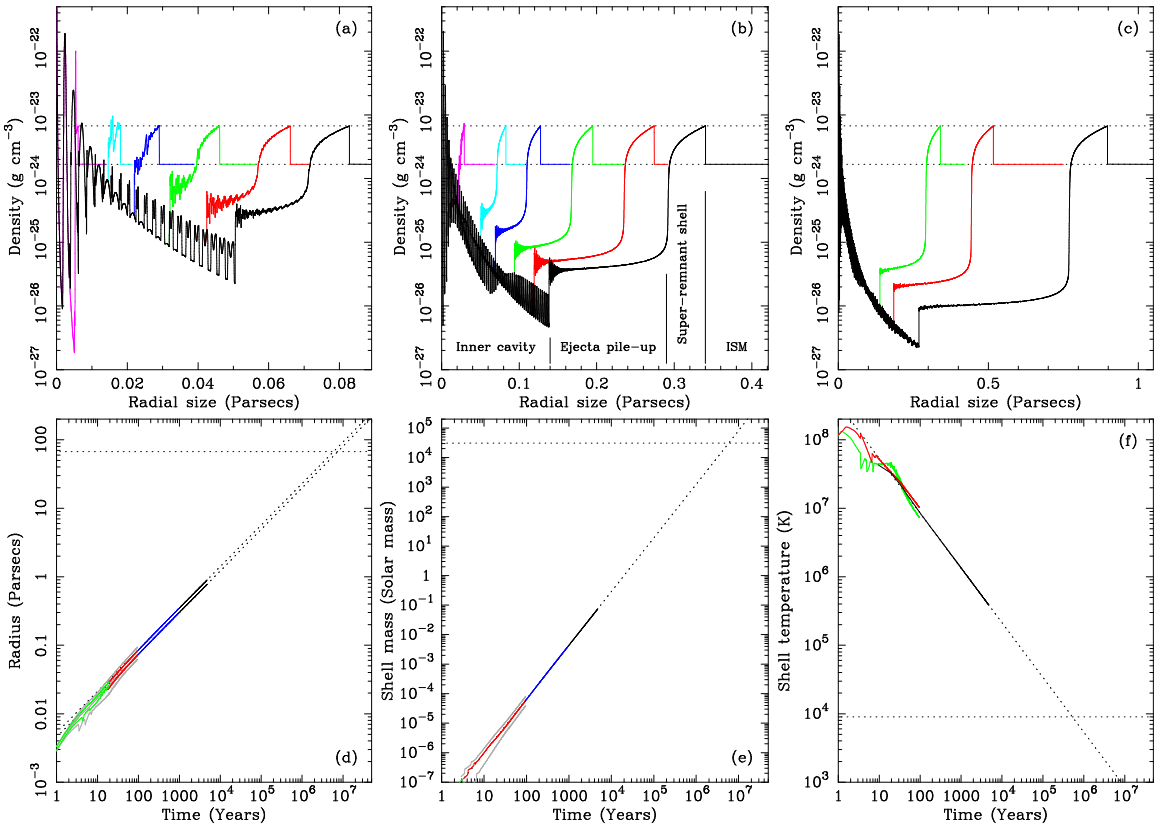

## Methods

1. M31N 2008-12a

2. Ground-based imaging observations

3. Hubble Space Telescope observations

4. Ground-based spectroscopic observations

5. The shell mass, luminosity, and motion of M31N 2008-12a

6. Hydrodynamic modelling

## Methods

### 1. M31N 2008-12a

The recurrent nova M31N 2008-12a is located in the outer disk in the north-eastern part of M31 with equatorial coordinates $0^h45^m28.89^s$ +41°54′10.2″ (J2000)[27]. Eruptions have been detected in each year from 2008 to 2016 and recovered from archival X-ray observations taken in 1992, 1993, and 2001[28]. The 2013-2016 eruptions have been studied extensively in the X-ray, ultraviolet, and optical. The recurrence period of the system is 347±10 days[11], although an alias of 174±10 days[29] cannot yet be excluded. The mass donor has been identified as a 'red clump' star[14], but based on Galactic systems[6] is most likely to be a low-luminosity red giant, with accretion driven either by Roche lobe overflow or by the red giant wind. Spectroscopy of the 2012-2016 eruptions has shown strong evidence for the deceleration of the 12a ejecta over the first 5 days post-eruption as they interact with circumbinary material, which must be replenished between each eruption[11,18]. The most likely source of this material is from a donor wind, although an accretion disk wind has also been proposed[14].

Short recurrence times are driven by the combination of a high-mass white dwarf and a high mass accretion rate[4]. Among Galactic systems, U Scorpii exhibits the shortest recurrence period of ~10 years[6], although recently a number of other short-period (<10 years) systems have been discovered in M31[30].

### 2. Ground-based imaging observations

Nebulosity in the region around 12a had first been identified as a 'ring' like structure as part of a narrow-band survey of M31 undertaken in 1987[19], 21 years before the first optical eruption was discovered. Following the 2015 eruption of 12a, an inspection of Hα data collected using the Steward 2.3m Bok Telescope in 2005 and 2006 (see Fig. M1) marked the 'rediscovery' of the nebulosity but its first association with the recurrent nova[18]. A further series of 20×180s narrow-band Hα images of the nebula were taken using the IO:O CCD camera on the 2.0m fully robotic Liverpool Telescope (LT[31]) on 2014 July 30[18]. The narrow-band Hα filter on the LT has a bandpass full-width at half-maximum of 10nm and therefore contains Hα+[N II] emission. Those data were supplemented by an additional series of 20×180s Hα+[N II] images taken between the 2014 and 2015 eruptions of 12a and by 66×300s Hα+[N II] images obtained during the 2014 eruption[18]. When combined, these produced a new deep 27ks Hα+[N II] image of 12a and the surrounding region. A deep (55.6ks) broad-band Sloan $r'$-band image of the region was also produced by combining all available LT observations of the 2014 and 2015 eruption campaigns[11,18].

These data were processed, co-aligned, and stacked using standard tools with the IRAF[32] environment. DAOPHOT[33] was then utilised to perform photometry on all sources common to the Hα and $r'$ images, which were then removed, before photometrically aligning the data, and subtracting. The resultant continuum-subtracted Hα+[N II] image is shown in Fig. 1a.

**3. Hubble Space Telescope observations**

Ten orbits of Cycle 24 *Hubble Space Telescope* time were used to obtain Hα+[N II] imaging of the nebulosity around 12a (proposal ID: 14651). These observations were conducted on 2016 December 7, 8, 9, 10, 11, and 17. By chance, the 2016 eruption of 12a was captured in those observations on December 17[34].

The orbits were split into five pairs, with each pair collecting narrow-band imaging through the F675N (Hα+[N II]) filter for one orbit, and through the F645N filter (for continuum subtraction) for the second orbit; both utilising the UVIS mode of the Wide Field Camera 3 instrument. The total exposure time for each F675N and F645N orbit was 2694s and 2805s, respectively.

For each filter, a three-point dither was applied to enable the removal of detector defects and cosmic ray rejection. The dither pattern of each visit was offset to allow for enhanced spatial resolution. A post-flash signal of 12 electrons was included to minimise losses due to charge transfer efficiency. These data were reduced using the STScI calwfc pipeline[35], and Drizzlpac was used to align and combine to create the images (Fig. 1b & 1c) with spatial resolution of 0.0333″/pixel (~0.1 parsecs at the distance of M31). Continuum subtraction was achieved by first flux aligning the images using resolved stellar sources in both the F645N and F675N images.

*Hubble Space Telescope* Wide Field Camera 3 imaging of the region around 12a were taken through broad-band optical and ultraviolet filters during cycle 23[(14)] (proposal ID: 14125). These images were processed and combined in a similar fashion to the narrow-band Hα+[N II] images. A sample of these images are shown in Fig. M2. Consistent with the ground-based spectra, there is no evidence for any continuum emission from any part of the super-remnant in these images.

**4. Ground-based spectroscopic observations**

Five 30 minute spectra spanning the wavelength range from 367.0 to 787.0nm were taken with the OSIRIS instrument on the Gran Telescopio Canarias on 2017 January 16. We used a slit width of 0.6″ orientated east-west and the 'R1000B' grism, achieving a spectral resolution of 0.53nm. After image reduction, cosmic ray, and two-dimensional sky background removal using IRAF, the five spectra were co-added, and the one-dimensional spectra of the super-remnant shell and eastern ‘knot’ were optimally extracted using PAMELA (part of the Starlink software package[36]).

The spectrum of the super-remnant outer shell is presented in Fig. 2; in Fig. M3 we present the spectrum of an inner eastern knot. In addition to hydrogen Balmer series lines and He I (587.6nm), the spectrum of the outer shell included emission lines from the resolved [N II] (654.8/658.4nm) and [S II] (671.6/673.1nm) doublets, the unresolved [O II] (372.6/372.9nm) doublet can be seen, all on top of a negligible continuum flux. There is little evidence for any other species, including the forbidden lines of O III. While the ‘knot’ spectrum is broadly similar to that of the super-remnant shell, no helium lines are detected and the [O II] emission in the knot has been joined by [O III] (495.9/500.7nm) emission. Although the [O III] (500.7nm) line intensity surpasses that of $H\beta$ and the [N II] (658.4nm) line is as strong as $H\alpha$, the [O III] (436.3nm) line and again the [N II] (575.5nm) line are not present. The [N II]/$H\alpha$ ratio in the knot is $1.23\pm0.08$, and [S II]/$H\alpha = 1.54\pm0.09$, the [S II] doublet ratio ([S II] 6717/[S II] 6731) is $1.41\pm0.07$, and the [O III] (500.7nm)/$H\beta$ ratio is $1.8\pm0.5$. The line ratios in the knot spectrum indicate a similar density upper limit to the super-remnant shell. However, the line ratios also point toward the inner knot containing much more strongly ionised gas.

**5. The shell mass, luminosity, and the motion of M31N 2008-12a**

The upper limit on the shell mass of the super-remnant was estimated from the imaging and spectroscopy. Based on studies of Galactic nova shells[37] we assumed a bi-axial geometry (either prolate or oblate). We used the projected semi-major and semi-minor axes of 67 and 45pc, respectively, a shell thickness of 22%, and the [S II] electron density upper limit of $100 cm^{-3}$. We note that the measured shell thickness ratio is invariant to projection effects. We derive shell mass upper limits of $7\times10^5$ Msun and $10^6$ Msun for prolate and oblate geometries, respectively. We also note that the [S II] doublet is not a sensitive probe of densities below $100 cm^{-3}$ (ref. 22).

Using the *Hubble Space Telescope* continuum-subtracted H$\alpha$+[N II] image (Fig. 1b) we computed the integrated H$\alpha$+[N II] flux from the super-remnant to be $7\times10^{-17}$ W $m^{-2}$. When accounting for the distance to M31 of 770±19 kpc[38], we find that the total H$\alpha$+[N II] luminosity of the super-remnant is 1300±200 solar luminosities.

In Fig. 1 we show that 12a is offset from the geometric centre of the super-remnant by 13 parsecs. The geometric centre was determined by the best-fitting ellipse to the optical imaging (see Fig. 1 left panel). To attain such a displacement over $6\times10^6$ years a transverse velocity of 2.1 km/s is required. All the spectra of 12a in eruption show no evidence for a significantly radial component to the system velocity[11,15,18,27,34].

**6. Hydrodynamic modelling**

The hydrodynamic simulations were performed with the Morpheus program, an MPI-OpenMP Eulerian second-order Godunov simulation code with options of Cartesian, spherical and cylindrical coordinates which includes radiative cooling and gravity. Morpheus combines well-established 1D (Asphere[39]), 2D (Novarot[40]) and 3D (CubeMPI[41]) codes written by the Manchester-Liverpool astrophysics groups into a single framework. For the purposes of these simulations we assumed one-dimensional spherical symmetry.

Based on observations of 12a and theoretical modelling of the eruptions[11,13,14,18], we assumed the following model for the system: The mass donor is a red giant with a wind mass loss rate (after any accretion on to the white dwarf) of $2.6\times10^{-8}$ Msun/year, the terminal velocity of the red giant wind is 20 km/s, and this wind blows continuously. The mass loss from the white dwarf, via nova eruptions, is modelled as a wind with a constant mass-loss rate and velocity which has a simple top-hat function 'on' for 7 consecutive days in every 350 days (the nova recurrence period). The total mass ejected by each nova eruption is $5\times10^{-8}$ Msun, the ejecta have a terminal velocity of 3,000 km/s. By injecting mass with terminal velocity we can neglect gravity. As the spatial resolution of the larger simulations is smaller than the expected orbital separation, both the donor star and the white dwarf are assumed to be spatially coincident at the origin. Mass injection via a wind or nova ejecta is effected by means of a boundary condition at the inner boundary of the simulation grid. The energy of the injected mass is dominated by kinetic energy. The grid is uniformly spaced and the maximum domain size is predetermined to contain the outer edge of the super-remnant. For computational efficiency, the domain is actively resized as the super-remnant grows by the addition of new cells.

Simulations were conducted as follows: An initial run, following 20 eruptions, with a spatial resolution of 0.02AU per cell and maximum domain size $10^{17}$cm (6667AU) was conducted to 'bench-mark' the lower-resolution simulations. This was followed by a simulation of 100 eruptions with 0.2AU/cell and maximum domain $3\times10^{17}$cm, 1,000 eruptions at 0.4AU/cell and maximum domain $1.2\times10^{18}$cm, and finally 5,000 eruptions with a resolution of 1AU/cell and maximum domain $4.4\times10^{18}$cm (1.43pc). All of these simulations prepopulated the domain with a low pressure, cold (90K), uniform ISM density of 1 hydrogen atom per $cm^3$, and did not invoke radiative cooling. Fig. M4 illustrates the consistency between the models at all resolutions particularly within the outer dense shell of the super-remnant.

Simulations of 100 eruptions were also conducted, as above, to explore the effect of different ISM densities on the super-remnant. This effect is shown in Fig. 3d. As any energy lost due to radiative cooling can affect the dynamics of a system, a simulation of 1,000 eruptions, again as above, was also conducted while utilising the radiative cooling module of Morpheus with a suitable cooling curve[39]. The results of this simulation are presented in Fig. M5 and are compared to the uncooled version. Again, although the details of the freely expanding nova ejecta are altered, the gross-structure of the super-remnant shell is consistent. We also note that radiative cooling is inefficient above $10^{6.5}$K, which is the case for all the material in the remnant shell. Therefore, for computational ease, we moved on to simulate greater numbers of ejecta while ignoring radiative cooling effects. Fig. 2f shows that eventually the super-remnant shell will cool sufficiently that radiative cooling may be important, however, the efficiency of such cooling also depends on the square of the density, so we do not expect cooling to greatly affect the results at later times.

The remnant mass (Fig. 3e) was computed by integrating over the super-remnant shell, defined as the outer region whose density is above that of the ISM (see Fig. 3b). Extrapolation of the simulations to greater time-scales was performed by fitting the super-remnant growth curves, as shown in Fig. 3d & 3e, with modified power-laws of the form: $\ln f(r) = a + b\,( c + \log r )^d$, with $c = 0$; in all cases $d$ was sufficiently close to unity that it could be ignored.

**Methods Figure 1.**

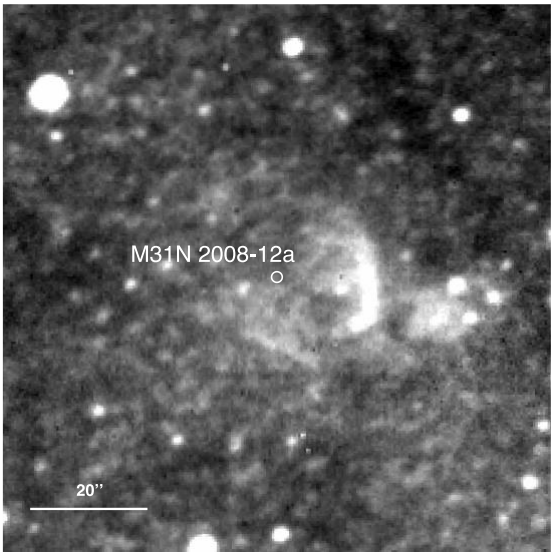

**Methods Figure 2.**

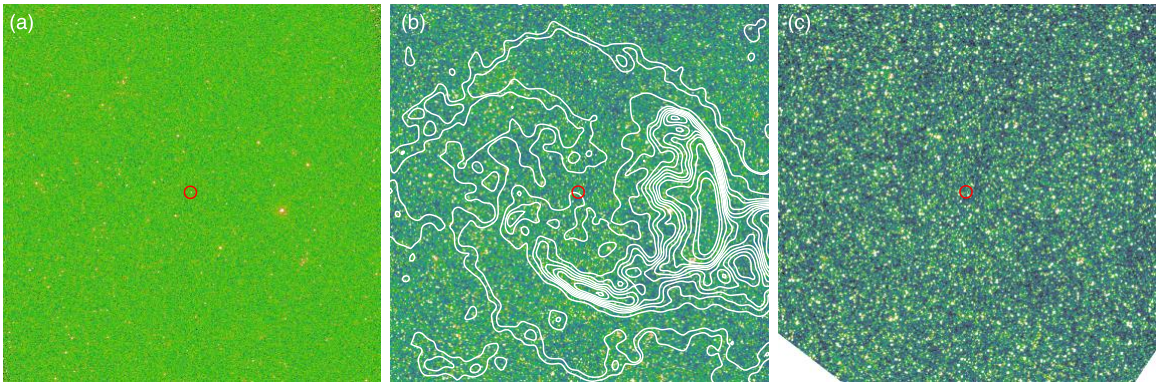

**Methods Figure 3.**

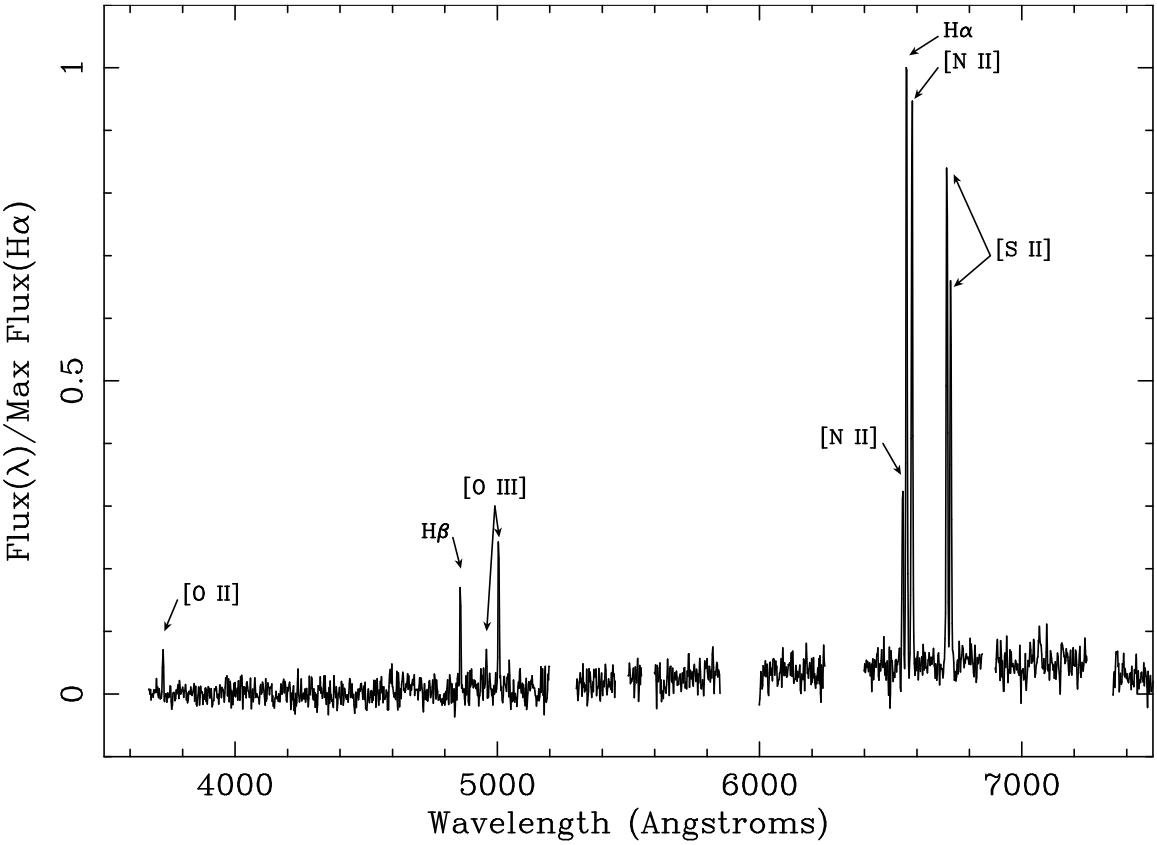

**Methods Figure 4.**

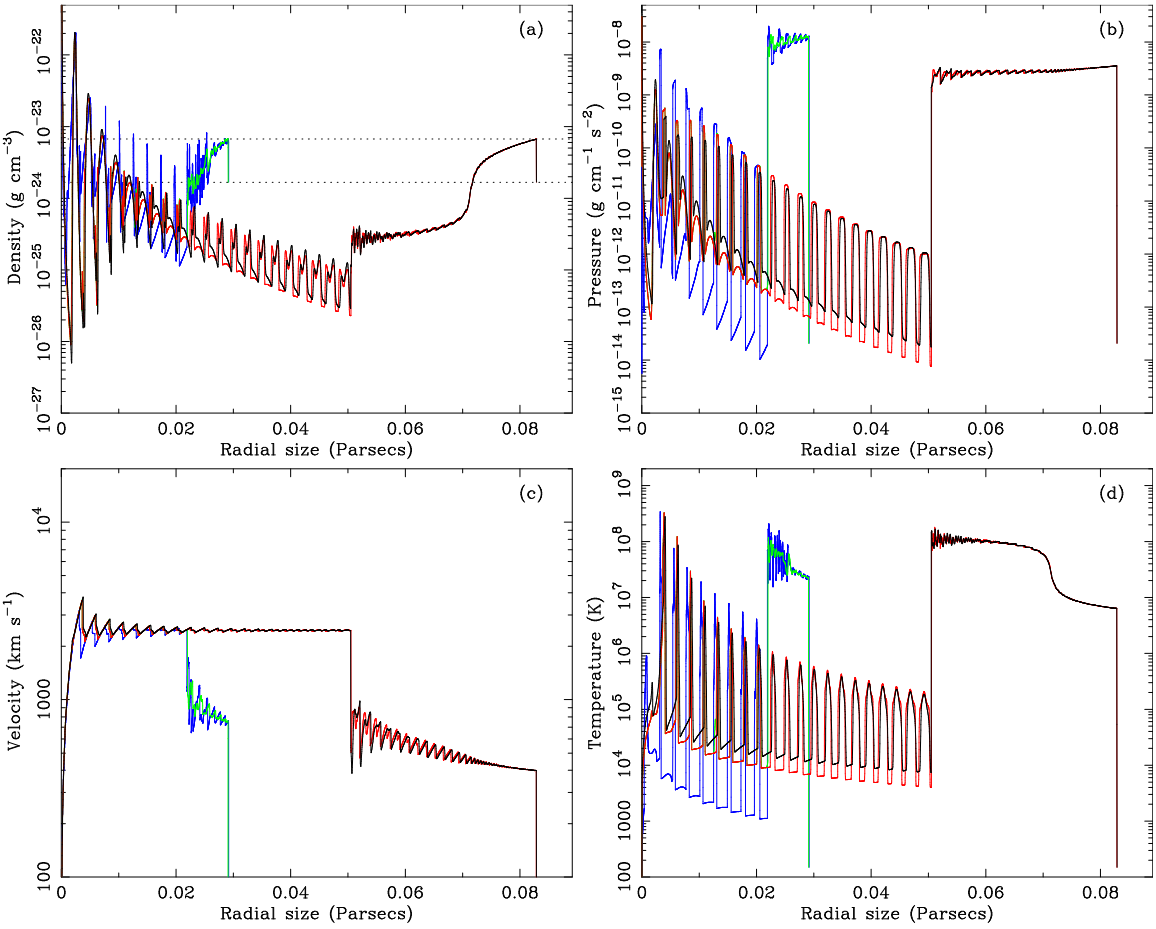

**Methods Figure 5.**

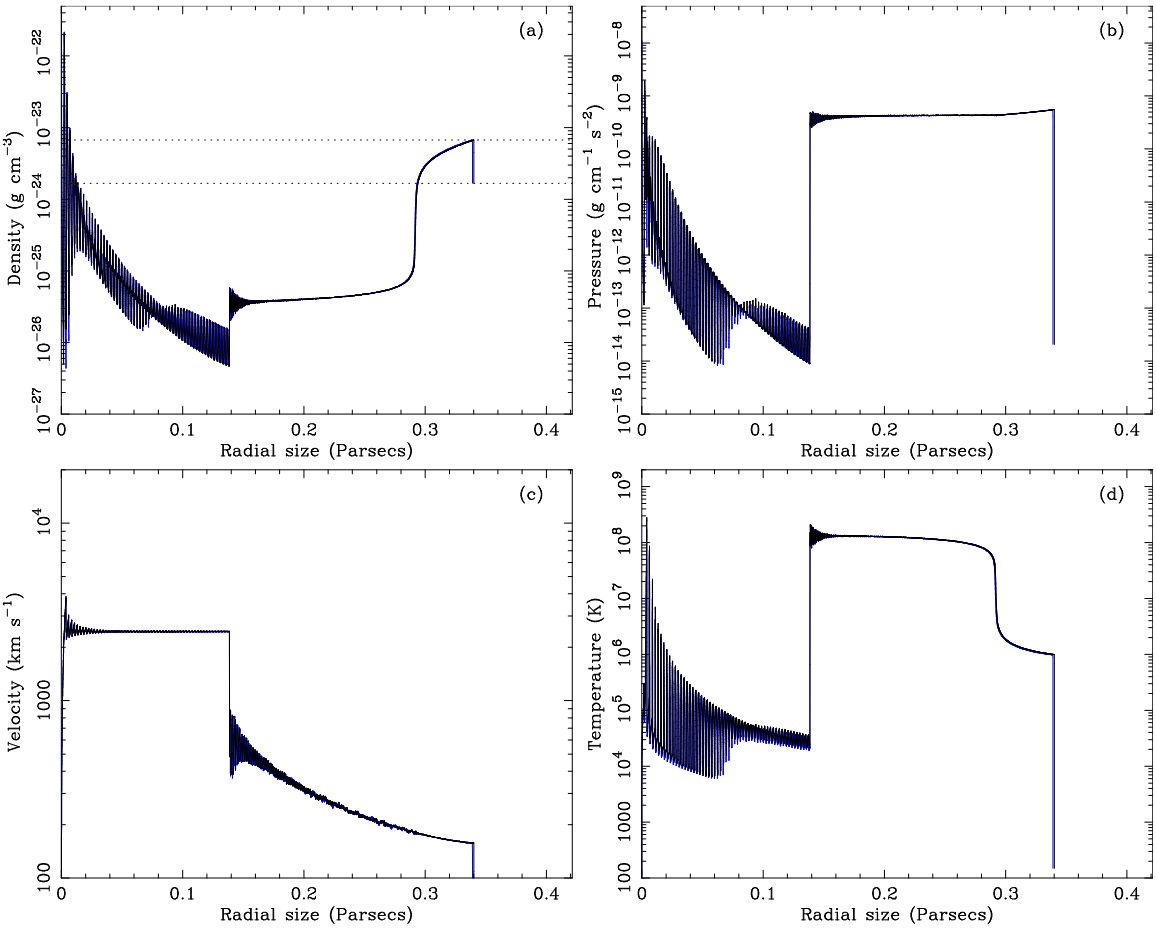

**Methods references**